\documentclass[lettersize,journal]{IEEEtran}
\usepackage{amsmath,amsfonts}
\usepackage{color}
\usepackage{array}
\usepackage[caption=false,font=normalsize,labelfont=sf,textfont=sf]{subfig}
\usepackage{textcomp}
\usepackage{stfloats}
\usepackage{url}
\usepackage{verbatim}
\usepackage{graphicx}
\usepackage{cite}
\usepackage{booktabs}
\usepackage{makecell}
\usepackage{nomencl}
\usepackage{multirow}

\makeatletter  
\newif\if@restonecol  
\makeatother

\usepackage[linesnumbered,ruled,lined]{algorithm2e}
\usepackage{algpseudocode}

\SetKwFor{For}{for}{do}{endfor}

\newtheorem{remark}{\textbf{Remark}}

\usepackage{graphicx}

\begin{document}



\title{Defense Hardening of Electric-Hydrogen Distribution Networks against Extreme Weather: A Safety Guarded Approach with Endogenous Uncertainties}

\title{Defense Hardening of Electric-Hydrogen Networks against Extreme Weather: A Safety Guarded Approach with Endogenous Uncertainties}

\title{Defense Hardening of Electric-Hydrogen Networks against Extreme Weather: A Safety Guarded Approach with Endogenous Uncertainties}


\title{Defense Hardening of Electricity-Hydrogen Networks for Extreme Weather: A Risk Restriction Approach with Endogenous Uncertainties}

\title{Defense Hardening of Electricity-Hydrogen Networks for Extreme Weather: A Risk Control Approach with Endogenous Uncertainties}

\title{Defense Hardening of Electricity-Hydrogen Networks for Extreme Weather: A Risk Limitation Approach with Endogenous Uncertainties}

\title{Optimal Hardening Strategy for Electricity -Hydrogen Networks with Hydrogen Leakage Risk Control against Extreme Weather}

\author{
	\IEEEauthorblockN{Sicheng Liu,~\IEEEmembership{Student Member,~IEEE,}}
	\IEEEauthorblockN{Bo Yang,~\IEEEmembership{Senior Member,~IEEE,}}
	\IEEEauthorblockN{Xin Li,}
	\IEEEauthorblockN{Xu Yang,~\IEEEmembership{Student Member,~IEEE,}}\\
	\IEEEauthorblockN{Zhaojian Wang,~\IEEEmembership{Member,~IEEE,}}
	\IEEEauthorblockN{Dafeng Zhu,~\IEEEmembership{Member,~IEEE,}}
	\IEEEauthorblockN{Xinping Guan,~\IEEEmembership{Fellow,~IEEE,}}
}


\maketitle

\begin{abstract}

Defense hardening can effectively enhance the resilience of distribution networks against extreme weather disasters. 
Currently, most existing hardening strategies focus on reducing load shedding. 
However, for electricity-hydrogen distribution networks (EHDNs), the leakage risk of hydrogen should be controlled to avoid severe incidents such as explosions.
To this end, this paper proposes an optimal hardening strategy for EHDNs under extreme weather, aiming to minimize load shedding while limiting the leakage risk of hydrogen pipelines. 
Specifically, modified failure uncertainty models for power lines and hydrogen pipelines are developed. 
These models characterize not only the effect of hardening, referred to as decision-dependent uncertainties (DDUs), but also the influence of disaster intensity correlations on failure probability distributions.
Subsequently, a hardening decision framework is established, based on the two-stage distributionally robust optimization incorporating a hydrogen leakage chance constraint (HLCC).
To enhance the computational efficiency of HLCC under discrete DDUs, an efficient second-order-cone transformation is introduced. 
Moreover, to address the intractable inverse of the second-order moment under DDUs, lifted variables are adopted to refine the main-cross moments.
These reformulate the hardening problem as a two-stage mixed-integer second-order-cone programming, and finally solved by the column-and-constraint generation algorithm. 
Case studies demonstrate the effectiveness and superiority of the proposed method.

\end{abstract}

\begin{IEEEkeywords}
Electricity-hydrogen distribution network, resilience enhancement, defense hardening, decision-dependent uncertainties, risk control.
\end{IEEEkeywords}

\addcontentsline{toc}{section}{Nomenclature}
\section*{Nomenclature}
\small
\subsection{Indices}
\begin{IEEEdescription}[\IEEEusemathlabelsep\IEEEsetlabelwidth{$G_{m}^{P2H,\max}G_{m}^{H2}$}] 
\item[$i,j,h$] Index of grid node.
\item[$m,n,o$] Index of hydrogen node.
\item[$ij$] Index of power line.
\item[$mn$] Index of hydrogen pipeline.
\item[$t$] Index of time period.
\item[$zn$] Index of zones in distribution networks.
\end{IEEEdescription}
\vspace{-0.5cm}
\subsection{Sets}
\begin{IEEEdescription}[\IEEEusemathlabelsep\IEEEsetlabelwidth{$G_{m}^{P2H,\max}G_{m}^{P2}$}] 
\item[$\mathcal{L}$] Set of power lines.

\item[$\mathcal{P}$] Set of hydrogen pipelines.
\item[$\mathcal{T}$] Set of moments $t$.

\item[$\mathcal{Z}$] Set of zones.
\item[$\mathcal{P}^{\mathrm{SSA}}$] Set of hydrogen pipelines in the safety-sensitive area.
\item[$\mathcal{S}^\mathrm{H}$] Set of hydrogen energy storages.
\item[$\mathcal{N}^\mathrm{E}$] Set of grid nodes.
\item[$\mathcal{N}^\mathrm{H}$] Set of hydrogen distribution network nodes.
\item[$\mathcal{P}_v$] Ambiguity set of wind speed probabilities.
\item[$\mathcal{P}_{a}$] Ambiguity set of components' failure probabilities.
\item[$\mathcal{P}_{a}^\mathrm{L}$] Lifted partial cross-moments ambiguity set of component failure probabilities.
\item[$\mathcal{A}_v$] Support set of wind speed.
\item[$\mathcal{A}_{a}$] Support set of components' failure.
\item[$\mathcal{A}_{a}^\mathrm{L}$] Support set of components' failure with lifted variables.
\item[$\delta(i),\delta(n)$] Set of superior nodes of node $i$/$n$ in the grid/hydrogen network.
\item[$\pi(i),\pi(n)$] Set of subordinate nodes of node $i$/$n$ in the grid/hydrogen network.
\end{IEEEdescription}
\vspace{-0.5cm}
\subsection{Variables}
\begin{IEEEdescription}[\IEEEusemathlabelsep\IEEEsetlabelwidth{$G_{m}^{P2H,\max}G_{m}^{H2}$}] 
\item[$v_{ij,t}$] Wind speed suffered by power line $ij$ at $t$.
\item[$r_{ij,t}$] Rainfalls suffered by hydrogen pipeline $mn$ at $t$.
\item[$f_{i,t}^{(*)}$] Failure probability of component $(*)$ with index $i$ at $t$.
\item[$x_{ij/mn}^{\mathrm{H}}$] Binary variable indicating whether the power line $ij/mn$ is hardened or not.
\item[$a^\mathrm{l}_{ij,t}$] Binary variable for whether the power line $ij$ is destroyed at $t$.
\item[$a^\mathrm{p}_{mn,t}$] Binary variable for whether the hydrogen pipeline $mn$ is destroyed at $t$.
\item[$u^\mathrm{l}_{ij,t}$] Binary variable for the $ij$ power line's state at $t$: destroyed or not.
\item[$u^\mathrm{p}_{mn,t}$] Binary variable for the $mn$ hydrogen pipeline's state at $t$: destroyed or not.
\item[$x_n^{\mathrm{E_0}}$] Continuous variable indicating the initial hydrogen storage allocation of node $n$.
\item[$E_n,t$] Amount of hydrogen storage of node $n$ at $t$.
\item[$P_{i,t}^{\mathrm{ls}},Q_{i,,t}^{\mathrm{ls}}$] Active/Reactive power load shedding of node $i$ at $t$.
\item[$G_{m,t}^{\mathrm{ls}}$] Hydrogen load shedding of node $i$ at $t$.
\item[$P_{ij,t},Q_{ij,,t}$] Active/reactive power of power line $ij$ at $t$.
\item[$G_{mn,t}$] Circulation of hydrogen pipeline $mn$ at $t$.
\item[$V_{i,t}^{\mathrm{sqr}}$] Square of voltage at grid node i at t.
\item[$P_{i,t}^{\mathrm{H2P}},Q_{i,t}^{\mathrm{H2P}}$] Active/reactive power generated by H2P fuel cell at node $i$ at $t$.
\item[$P_{i,t}^{\mathrm{P2H}},Q_{i,t}^{\mathrm{P2H}}$] Active/reactive power consumed by P2H electrolyzer at node $i$ at $t$.
\item[$P_{i,t}^{\mathrm{sub}},Q_{i,t}^{\mathrm{sub}}$] Active/reactive power supplied by substation at node $i$ at $t$.
\item[$F_{n,t}^{\mathrm{H2P}},F_{n,t}^{\mathrm{P2H}}$] Hydrogen consumed/generated by H2P/P2H at node $n$ at $t$.
\item[$F_{n,t}^{\mathrm{s+}},F_{n,t}^{\mathrm{s-}}$] Charging and discharging of hydrogen storage at node $n$ at $t$.
\item[$F_{n,t}^{\mathrm{UT}}$] Hydrogen supplied by upper transmission network at node $n$ at $t$.
\end{IEEEdescription}
\vspace{-0.5cm}
\subsection{Parameters and Constants}
\begin{IEEEdescription}[\IEEEusemathlabelsep\IEEEsetlabelwidth{$G_{m}^{P2H,\max}G_{m}^{H2}$}] 
\item[$(\cdot)^{\mathrm{max}},(\cdot)^{\mathrm{min}}$] Maximum and minimum limits for item $(\cdot)$ .
\item[$c_{ij}^\mathrm{H}/c_{mn}^\mathrm{H}$] Harden cost for power lines $ij$ or hydrogen pipelines $mn$.
\item[$C^{\mathrm{H}}$] Maximum harden budget.
\item[$E_0$] Amount of the pre-disaster allocable hydrogen storage.
\item[$K_{\mathrm{CC}}$] Maximum tolerable amount of damage to safety-sensitive areas.
\item[$c_{i}^{\mathrm{e,ls}}$] Cost of power load shedding at node $i$ per unit.
\item[$c_{n}^{\mathrm{h,ls}}$] Cost of hydrogen load shedding at node $n$ per unit.
\item[$R_{ij},X_{ij}$] Impedance/conductance of power line $ij$.
\item[$P_{i,t}^\mathrm{L},Q_{i,t}^\mathrm{L}$] Active/reactive power load of node $i$ at $t$.
\\
\item[$G_{n,t}^\mathrm{L}$] Hydrogen load of node $n$ at $t$.
\\
\item[$\eta ^{\mathrm{s+}},\eta ^{\mathrm{s-}}$] Hydrogen storage charging and discharging energy efficiency.
\item[$\beta^{\mathrm{H2P}},\beta^{\mathrm{P2H}}$] Energy conversion efficiency of P2H and H2P.
\end{IEEEdescription}

\normalsize
\section{Introduction}

\IEEEPARstart{W}{ith} growing environmental challenges and the finite nature of fossil fuels, electricity-hydrogen coupling systems are attracting considerable attention \cite{aliaksei2025Green}.
In particular, a highly integrated electricity-hydrogen distribution network (EHDN) enables flexible operations through distributed energy storage or conversion at hydrogen stations and hydrogen transport via pipelines.

Despite their advantages, the operation of EHDNs is threatened by the increasing frequency of extreme weather events, such as hurricanes, floods, and hailstorms. These disasters can cause significant damage to the energy infrastructure, including power lines or hydrogen pipelines \cite{Executive2013_back_new4,tianwen2024Hydrogen}.

As a result, enhancing the resilience of energy systems has gained considerable attention, which primarily focuses on reducing load shedding costs \cite{2022Back06}. 
Nonetheless, given that hydrogen is both flammable and explosive, any leakage poses significant risks, particularly in sensitive areas. 
Therefore, the leakage risk of hydrogen pipelines in EHDNs must be controlled under extreme weather conditions. 


Fortunately, defensive hardening can effectively enhance the reliability of power lines and hydrogen pipelines \cite{chuan2018Back05}. 
However, for EHDNs, two key challenges remain:
(1) Accurately characterizing the failure uncertainties of power lines and hydrogen pipelines, which are influenced by both hardening decisions and disaster intensity. This is fundamental to ensuring the optimality of hardening strategies.
(2) Effectively controlling the leakage risk of hydrogen pipelines through hardening, which is crucial for the safety of EHDNs and their broader adoption.



%
%
%

As for the characterization of failures, early research assumed that lines would not fail after hardening \cite{shanshan2018RO01}, or used a polyhedral set to represent the uncertainty of failures \cite{wei2016RO02,xiang2024DDURO01Resilient}, modeling it as a robust optimization (RO) problem. 
To more accurately characterize the effect of hardening, decision-dependent uncertainties (DDU) were introduced, where the failure probabilities depend on whether the lines are hardened or not \cite{shen2024CPSRenXing}.
Then, to make optimal hardening decisions under the above DDUs, stochastic programming (SP) and distributional robust optimization (DRO) were employed, which typically construct a two-stage model to represent pre-disaster hardening and post-disaster energy dispatch.
Ma et al. \cite{shanshan2018DDUSP01} and Zhang et al. \cite{weixin2023DDUSP02} used SP with DDU for hardening decisions at the distribution and transmission network levels, respectively.
Furthermore, the DRO method under DDU was developed, which considers resilience under the worst-case uncertainty distribution of failure probabilities.
Based on DRO, an earthquake-oriented resilience enhancement method was proposed through hardening measures and transmission expansions \cite{diego2023DDUDRO01}.
Incorporating the uncertainty of typhoon trajectories, Li et al. \cite{yujia2023DDUDRO02} developed a scenario-based distributionally robust hardening strategy.
However, limited by DDUs, these works overlooked the correlations of disaster intensity and their impact on failures.  
Although the second-order moment can introduce correlations to DRO \cite{erick2010DROXuangamma1gamma2}, 
DDUs make the inverse of second-order moments non-analytically expressible, making it difficult to solve.
This oversight of correlations may lead to a more conservative representation of uncertainties, thereby reducing the effectiveness of hardening under budget constraints.
%


Moreover, for risk control related to hydrogen leakage, chance constraints 
can limit the probability of constraint violations under uncertainties. 
In the context of resilience enhancement,
Chu et al. \cite{zhongda2021CC03} proposed a resilience scheduling method for microgrids, utilizing chance constraints to ensure power frequency stability under disasters.
Additionally, a decentralized dispatch method was proposed considering communication outages of the energy system, where chance constraints were used to restrict the energy balance considering uncertainties in renewable energy generation \cite{tong2022CC02}.
These studies used chance constraints to restrict the continuous variables in energy dispatch.
However, for the leakage risk control, the uncertain states of hydrogen pipelines are discrete and decision-dependent.
This property makes the commonly used solution methods based on conditional value-at-risk and duality \cite{zhichao2019DROCC} intractable due to the curse of dimensionality.

Above all, in response to the threat of extreme weather events,
this paper proposes an optimal hardening decision strategy for EHDNs,
aiming to minimize load-shedding costs while limiting the leakage risk of hydrogen pipelines.


In summary, the contributions of this paper are as follows:

\begin{enumerate}
	\item{
		
		
		
		
		To more accurately characterize the failure probabilities of power lines and pipelines under extreme weather events, a modified DDU model was proposed using fragility analysis. The failure uncertainty model considers not only the effect of hardening but also the influence of disaster intensity and its correlations on failure probability distributions.
		
	}
	\item{

	To obtain the optimal hardening strategy based on the failure uncertainties,
a two-stage distributionally robust decision framework is proposed, incorporating a hydrogen leakage chance constraint (HLCC).
The framework aims to minimize load shedding while limiting the leakage risk of hydrogen pipelines in sensitive areas. 
Additionally, the HLCC also provides a criterion for the minimum hardening budgets under varying disaster intensities and desired risk levels. 
	
	
	}

	\item{
	
	Several customized transformation methods are proposed to overcome the difficulties in solving the problem.
	First, to address the intractability of HLCC under discrete DDUs, an efficient second-order-cone transformation is introduced.
	Furthermore, to deal with the non-analytical inverse of the second-order moment under DDUs, lifted variables are adopted to refine the main-cross moments. 
	Finally, the reformulated problem is solved using the column-and-constraint generation algorithm. 


}

\end{enumerate}

The remainder of this paper is organized as follows. 
Section II establishes the failure uncertainty model and formulates the hardening decision problem.
Section III presents the solution methodology, including the reformulation of the HLCC and the solution of the two-stage problem under DDUs.
Case studies are presented in Section IV.
Finally, Section V concludes the paper.


\section{Problem Formulation}
\subsection{Scenarios and Assumptions}

In the typical EHDN scenario \cite{sidun2022DianQingPeiWangChangJingOptimal}, distributed generators (DGs) and hydrogen stations are deployed distributed in the EHDN.
Hydrogen stations include hydrogen storage, electrolyzers, and fuel cells, coupling the two networks through energy conversions.
Electricity and hydrogen are transported via power lines and hydrogen pipelines.

Extreme weather events, such as typhoons, floods, and inundation, may damage power lines and hydrogen pipelines.
In densely populated or industrial areas, hydrogen leakage may cause severe incidents, necessitating stringent leakage risk control. 
In this paper, these areas are referred to as safety-sensitive areas (SSAs).
%
%

Without loss of generality, the following assumptions are made:
\begin{enumerate}
	\item{Considering the geographic characteristics, distribution networks can be divided into several zones, with the intensity of extreme weather in each zone assumed to be uniform \cite{haibo2021Back10}.}
	\item{
		Hydrogen equipment, such as electrolyzers, fuel cells, and hydrogen storage, is placed indoors and thus protected from extreme weather events. However, hydrogen pipelines remain vulnerable to waterlogging caused by heavy rainfall \cite{chuan2018Back05}.}

	
\end{enumerate}
\vspace{-10pt}
\subsection{Modeling Failure Uncertainties}
This subsection derives the failure uncertainty models for power lines and hydrogen pipelines.
The models are described in the form of a distributionally robust ambiguity set, which characterizes the statistical information of the failure probability distributions.
In this subsection, typhoons and the accompanying heavy rainfall are considered typical extreme weather events.


\subsubsection{The Uncertainty model of Disaster Intensity}

Considering that predictions of extreme weather events, such as typhoons and rainfall, are quite advanced \cite{robert2013Typhoon}, it is reasonable to assume that the probability distribution of disaster intensity is known. 
Regarding the correlations of intensity \cite{jonathan2011Typhoon}, the upcoming wind speed and rainfall intensity can be represented by a second-order moment ambiguity set as follow, which characterizes all possible probability distributions.

\begin{gather*}
	\mathcal{P} _{\boldsymbol{d}}\!:=\!\left\{ \mathbb{P}_{\boldsymbol{d}}\!\in\! \mathcal{M} _+(\mathcal{A}_{\boldsymbol{d}} )\left|\!\!\! \begin{array}{c}
		\left( \mathbb{E} _P[\boldsymbol{d}]-\bar{\boldsymbol{d}} \right) ^T \!\!Q_d ^{-1}\!\left( \mathbb{E} _P[\boldsymbol{d}]-\bar{\boldsymbol{d}} \right) \le \gamma_{{\boldsymbol{d}},1}
		\\
		\mathbb{E} _P\left[ (\boldsymbol{d}-\bar{\boldsymbol{d}})(\boldsymbol{d}-\bar{\boldsymbol{d}})^T \right] \preceq \gamma_{{\boldsymbol{d}},2}Q_{\boldsymbol{d}}\\
	\end{array} \right.\!\!\!\! \right\}, \tag{1a}
	\\ 
	\mathcal{A} _{\boldsymbol{d}}=\left\{ \boldsymbol{d} \left| \begin{array}{c}
		v_{zn,t}\in \left[ v_{zn,t}^{\min},v_{zn,t}^{\max} \right]\\
		r_{zn,t}\in \left[ r_{zn,t}^{\min},r_{zn,t}^{\max} \right]\\
	\end{array} \right. \right\},  \tag{1b}
\end{gather*}
where 
$\boldsymbol{d}=\left\{ v_{zn,t},r_{zn,t} , \forall (zn,t) \in (\mathcal{Z,T}) \right\}$, and $v_{zn,t}$ denotes the wind speed and $r_{zn,t}$ denotes rainfalls suffered in zone $zn$ at moment $t$.
$\mathbb{P}_{\boldsymbol{d}}$ represents the probability distribution of $\boldsymbol{d}$. 
$\mathcal{A} _{\boldsymbol{d}}$ is the support set of $\boldsymbol{d}$, and $\mathcal{M} _+(\mathcal{A}_{\boldsymbol{d}} )$ is the set of all distributions on the sigma-field of $\mathcal{A}_{\boldsymbol{d}}$.
$\bar{\boldsymbol{d}}$ and $Q_{\boldsymbol{d}}$ are the expectation and second-order moment of the intensity, respectively.
Ratio $\gamma_{{\boldsymbol{d}},1}$ and $\gamma_{{\boldsymbol{d}},2}$ characterize the statistical errors, which can be obtained using the method in \cite{erick2010DROXuangamma1gamma2}. 
The support set $\mathcal{A} _{\boldsymbol{d}}$ bounds the upper/lower bounds of the intensity, and $\mathcal{M} _+(\cdot)$ represents the set of all distributions with a given support set. 


\subsubsection{Fragility Curves of Power Lines}

Fragility curves, describing the relationship between failure probabilities and the suffered disaster intensity \cite{jeffreya.2021Fragility}.
During typhoons, wires and poles could be damaged by mechanical vibrations. The empirical fragility curves of a single segment of a power line and a single pole can be expressed by \cite{min2014Fragilitymodel01}
	\begin{align*}
		&f_{ij, k,t}^p\left(v_{ij,t},x_{ij}\right)\!=\!\min \left\{ a_{ij}(x_{ij})e^{b_{ij}(x_{ij})v_{ij,t} }, 1\right\},  \tag{2a}\\
		&f_{ij,l,t}^{w}\left( v_{ij,t},x_{ij} \right) \!=\!\max \left\{ f_{ij,l,t}^{w,d}\!\left( \!v_{ij,t}\! \right) \!,\chi _{ij,l}\left(\! x_{ij} \!\right)\! f_{ij,l}^{w,id}\!\left( v_{ij,t},x_{ij} \right) \!\right\}\!, \tag{2b}
	\end{align*}
	where subscripts $k$ and $l$ represent the index of wire segments and poles of power line $ij$. The segments of wires are divided by poles. $v_{ij,t}$ is the wind speed suffered by power line $ij$ at time $t$.
Binary variable $x_i^{\mathrm{H}}$ represents whether the component is hardened. 
$f_{ij,l}^{w,d}$ and $f_{ij,l}^{w,id}$ represent the failure rates directly blow down or indirectly struck by trees, respectively. $a_{ij}(x_{ij})$, $b_{ij}(x_{ij})$, and $\chi _{ij,l}(x_{ij})$ are the reliability parameters of the components, which can be determined using mechanistic analysis \cite{min2014Fragilitymodel01}.

Therefore, the failure probability for a power line $ij$ in the distribution network can be calculated using the following cumulative form, considering each line segment and pole in (2) \cite{yujia2023DDUDRO02}.
\begin{align*}
	f_{ij,t}^{\mathrm{L}}\left( v_{ij,t},x_{ij}^{\mathrm{H}} \right) =1-&\prod_{k=1}^{N_{ij}^{\mathrm{P}}}{\left( 1-f_{ij,k,t}^{p}\left( v_{ij,t},x_{ij}^{\mathrm{H}} \right) \right)}
	\\
	&\cdot \prod_{l=1}^{N_{ij}^{\mathrm{W}}}{\left( 1-f_{ij,l,t}^{w}\left( v_{ij,t},x_{ij}^{\mathrm{H}} \right) \right)},\tag{3}
\end{align*}
where $N^{P}_{ij}$ and $N^{W}_{ij}$ represent the number of poles and segments on power line $ij$.

\begin{figure}[!t]
	\centering
	\includegraphics[width=3.6in]{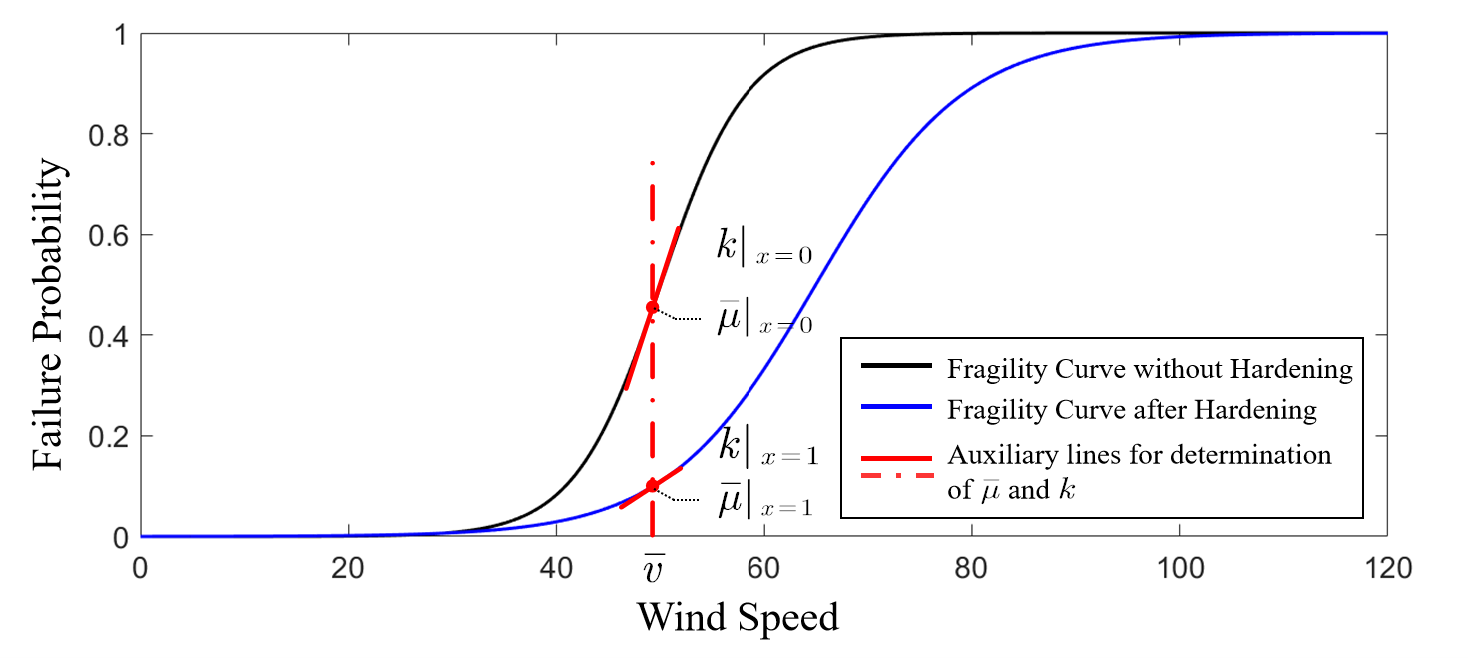}
	\caption{Illustration of fragility curves and related variables. 
		It can be observed that the failure probability still exists after hardening but is significantly reduced, which can be characterized as decision dependent uncertainty.
		$\bar{\mu}$ and $k$ are two parameters introduced in Section II-B.3), representing the probability of failure and the linearized slope of the fragility curve at the expected wind speed $\bar{v}$, respectively.}
	\label{fig_frag_curve}
	\vspace{-13pt}
\end{figure}

To more clearly illustrate the effect of hardening, the black and blue curves in Fig. \ref{fig_frag_curve} depict failure probabilities before and after hardening.
Note that the high-accuracy modeling of fragile curves is beyond the scope of this paper, since the subsequent DRO method does not rely on accurate fragile curves and can handle these inaccuracies.

\subsubsection{Fragility Curves of Hydrogen Pipelines}

	The fragility curve for a hydrogen pipeline segment can be expressed as 
	\begin{align*}
		f_{mn,0,t}^{\mathrm{Hy}}\!\left( \boldsymbol{r}_{mn},\boldsymbol{x}_{mn}^{\mathrm{H}} \right)\! =\!\phi \!\left( \frac{\ln \left( z_{mn,0}\left( \boldsymbol{x}_{mn}^{\mathrm{H}} \right) \sum_{\tau \le t}{r_{mn,\tau}} \right)}{\sigma _{mn,0}\left( \boldsymbol{x}_{mn}^{\mathrm{H}} \right)} \!\right),   \tag{4}
	\end{align*}
	where $\phi(\cdot)$ is the lognormal cumulative distribution function, which is well-suited for describing the failure probability of hydrogen pipelines \cite{yihuan2023Importance}.
	$r_{mn,t}$ is the rainfalls suffered by hydrogen pipeline $mn$ at time $t$, and the fragility is impacted by accumulated rainfall. $z_{mn,0}( \boldsymbol{x}_{mn}^{\mathrm{H}})$ and $\sigma _{mn,0}\left( \boldsymbol{x}_{mn}^{\mathrm{H}} \right)$ can be estimated by empirical data \cite{mohamed2018Reliability}.


The failure probability of hydrogen pipeline $mn$ in the distribution network can be calculated by
	\begin{align*}
		f_{mn,t}^{\mathrm{Hy}}\!\left(\boldsymbol{r}_{mn},x_{mn}^{\mathrm{H}} \right) =1\!-\!\prod_{k=1}^{N_{mn}^{\mathrm{Hy}}}{\!\left( 1\!-\!f_{mn,k,t}^{\mathrm{Hy}}\!\left( \boldsymbol{r}_{mn},x_{mn}^{\mathrm{H}} \right) \right)},
		\tag{5}
	\end{align*}
	where $N_{mn}^{\mathrm{Hy}}$ is the number of natural segments of the hydrogen pipeline $mn$.
	

\subsubsection{The Ambiguity Set of Failure Distributions}

The failure uncertainties of power lines and hydrogen pipelines are influenced not only by hardening but also by disaster intensity and its correlations.
To better characterize the uncertainties, a decision-dependent second-order moment ambiguity set is introduced, which describes all possible probability distributions of failures.
%
%
%
\begin{align*}
	\!\!\mathcal{P} _{\boldsymbol{a}}(\boldsymbol{x}^{\mathrm{H}})\!:=\!\left\{\!\!\!\! \begin{array}{c}
		\mathbb{P} _{\boldsymbol{a}}\in \mathcal{M} _+(\mathcal{A} _{\boldsymbol{a}}):\\
		\left( \mathbb{E} _P[\boldsymbol{a}]\!-\!\bar{\boldsymbol{\mu}}(\boldsymbol{x}^{\mathrm{H}}) \right) ^T\!\!Q\!\left( \boldsymbol{x}^{\mathrm{H}} \right) ^{-1}\!\!\left( \mathbb{E} _P[\boldsymbol{a}]\!-\!\bar{\boldsymbol{\mu}}(\boldsymbol{x}^{\mathrm{H}}) \right) \!\le \!\gamma\\
		\mathbb{E} _P\left[ (\boldsymbol{a}\!-\!\bar{\boldsymbol{\mu}}(\boldsymbol{x}^{\mathrm{H}}))(\boldsymbol{a}\!-\!\bar{\boldsymbol{\mu}}(\boldsymbol{x}^{\mathrm{H}}))^T \right] \!\preceq \!\gamma _2Q(\boldsymbol{x}^{\mathrm{H}})\\
	\end{array} \!\!\!\!\right\} 
	\tag{6}
\end{align*}
where $\boldsymbol{a}=\{a_{i,t}, \forall i \in \{\mathcal{L,P}\},\forall t \in \mathcal{T}\}$ is the binary random vector about the failures, in which $a_{i,t}=1$ represents component $i$ is destroyed at $t$, and conversely, $a_{i,t}=0$ indicates that it is not destroyed. 
$\mathbb{P}_{\boldsymbol{a}}$ represents the probability distribution of $\boldsymbol{a}$. 
 $\mathcal{A} _{\boldsymbol{a}}$ and $\mathcal{M} _+(\mathcal{A}_{\boldsymbol{a}} )$ is the support set and corresponding distributions set, respectively.
 $\bar{\boldsymbol{\mu}}(\boldsymbol{x}^{\mathrm{H}})$ and $Q(\boldsymbol{x}^{\mathrm{H}})$ are the expectation and second-order moment of $\boldsymbol{a}$, respectively.
 

To derive the analytic expression of (6),
the fragility curves are first linearly expanded to estimate the $\bar{\boldsymbol{\mu}}(\boldsymbol{x}^{\mathrm{H}})$ and $Q(\boldsymbol{x}^{\mathrm{H}})$ by $\widetilde{\boldsymbol{\mu}}(\boldsymbol{x}^{\mathrm{H}})$ and $\widetilde{Q}(\boldsymbol{x}^{\mathrm{H}})$.
Then, $\gamma_1$ and $\gamma_2$ are determined by a posteriori method to handle the error.

First, the linear expansion of the fragility curve $i$ at the expected wind speed $\bar{v}_i$ can be expressed as
\vspace{-3pt}
\begin{align*}
	\tilde{f}_i \,:=\widetilde{k_i} v+\widetilde{b_i},  \tag{7a}
\end{align*}
\vspace{-3pt}
where
\vspace{-3pt}
\begin{align*}
\widetilde{k_i}=k_{i}^{0}\left( 1-x_i^{\mathrm{H}} \right) +k_{i}^{1}x_i^{\mathrm{H}}, \tag{7b}
\\
\widetilde{b_i}=b_{i}^{0}\left( 1-x_i^{\mathrm{H}} \right) +b_{i}^{1}x_i^{\mathrm{H}}, \tag{7c}
\end{align*}
where $k_{i}^{0}/k_{i}^{1}$ and $b_{i}^{0}/b_{i}^{1}$ represent the slope and intercept of the linear approximation of the fragility curve $i$ at $\bar{v}_i$ with/without hardening, respectively.

So the affine approximation of $\bar{\mu}_i$ is
\vspace{-3pt}
\begin{align*}
\widetilde{\mu }_i\left( \boldsymbol{x}^{\mathrm{H}} \right) =\mathbb{E} _P\left[ \tilde{f}_i\left( v_i,x_i^{\mathrm{H}} \right) \right] =\tilde{f}_i\left( \bar{v}_i,x_i^{\mathrm{H}} \right) .
 \tag{8}
\end{align*}
\vspace{-3pt}
The estimate of the second-order moment can be derived by
\vspace{-1pt}
\begin{align*}
	&\widetilde{\mathbb{E}} _{Pa}\left[ (\boldsymbol{a}-\bar{\boldsymbol{\mu}}(\boldsymbol{x}^{\mathrm{H}}))(\boldsymbol{a}-\bar{\boldsymbol{\mu}}(\boldsymbol{x}^{\mathrm{H}}))^T \right]\\
	&:=\mathbb{E} _{Pv}\!\left[ \left( \widetilde{\boldsymbol{f}}\left( \boldsymbol{v},\boldsymbol{x}^{\mathrm{H}} \right)\! -\!\boldsymbol{f}\left( \overline{\boldsymbol{v}},\boldsymbol{x}^{\mathrm{H}} \right) \right) \left( \widetilde{\boldsymbol{f}}\left( \boldsymbol{v},\boldsymbol{x}^{\mathrm{H}} \right) \!-\!\boldsymbol{f}\left( \overline{\boldsymbol{v}},\boldsymbol{x}^{\mathrm{H}} \right) \right) ^T \right]\\
	&=\mathbb{E} _{Pv}\left[ \widetilde{k}\left( \boldsymbol{x}^{\mathrm{H}} \right) \cdot \left( \boldsymbol{v}-\overline{\boldsymbol{v}} \right) \left( \boldsymbol{v}-\overline{\boldsymbol{v}} \right) ^T\cdot \widetilde{k}\left( \boldsymbol{x}^{\mathrm{H}} \right) ^T \right]\\
	&\le \left( \widetilde{k}\left( \boldsymbol{x}^{\mathrm{H}} \right) \widetilde{k}\left( \boldsymbol{x}^{\mathrm{H}} \right) ^T \right) \odot Q_v:=\widetilde{Q}(x^{\mathrm{H}}), \tag{9}
\end{align*}
where $\odot$ represents the multiplication of the corresponding elements.

The error ratios $\gamma_1$ and $\gamma_2$ account for inaccuracies introduced by fragile modeling, linear approximation, and related factors.
These ratios can be determined through a posterior cross-validation method \cite{yiling2018CC_solve}.

Note that linearized model (7) is only used to estimate the second-order moment (9).
Even when the variant of disaster intensity is large, the overly relaxed second-order moment with $\gamma_{2}$ would not degrade the performance of the failure ambiguity set, compared to the existing first-order moment methods \cite{dimitris2019Adaptive}.

\vspace{-10pt}

\subsection{Formulation of the Hardening Decision Problem}

%


The hardening decision framework is formulated as a two-stage DRO model, aiming to determine optimal hardening decisions before in pre-disaster, while the second stage characterizes energy management and load shedding costs in post-disaster. In addition, the HLCC is introduced in the first stage to control hydrogen leakage risk.
The object can be expressed as 
\begin{align*}
	\underset{\boldsymbol{x}}{\min}\,&\underset{p_a\left( \boldsymbol{x} \right) \in \mathcal{P}_a}{\max}\,\,\mathbb{E}_{P}\left[ \underset{\boldsymbol{y}}{\min}\,\,g\left( \boldsymbol{x,y,a} \right) \right] .\tag{10}
\end{align*}

\subsubsection{First-Stage Pre-disaster Decisions with the HLCC}
In the pre-disaster, the hardening decisions and the allocation of hydrogen storage should meet budget and available capacity constraints.
\vspace{-2pt}
\begin{gather*}
	\sum_{ij\in \mathcal{L} \cup \mathcal{P}}{c_{ij}^{\mathrm{H}}x_{ij}^{\mathrm{H}}}\le C^{\mathrm{H}} ,\tag{11a}
	\\
	\sum_{n\in \mathcal{S}^\mathrm{H}}{x_{n}^{\mathrm{E}_0}}=E_0 ,\tag{11b}
	\\
	x_{n}^{\mathrm{E}_0}\leq E_{n}^{max} ,\tag{11c}
\end{gather*}
where $\boldsymbol{x}^{\mathrm{H}}=\{x_{ij}^{\mathrm{H}}, \forall ij \in \mathcal{L,P}\}$ and $\boldsymbol{x}^{\mathrm{E}_0}=\{x_{n}^{\mathrm{E}_0}, \forall n \in \mathcal{N}^\mathrm{H}\}$ represent hardening and hydrogen storage allocation in pre-disaster. $C^{\mathrm{H}}$ is the maximum hardening budget. 
$E_0$ is the total available hydrogen to be stored in pre-disaster.
$E_{i}^{\max}$ is the maximum capacity of energy storage $i$.


In addition, to control the hydrogen leakage risk in the SSA, the following HLCC is introduced.
\begin{align*}
	\underset{p_a\in \mathcal{P}_a}{\mathrm{inf}}\,\,\mathbb{P}\left( \sum_{t\in T}{\sum_{i\in \mathcal{P}^{\mathrm{SSA}}}{\left( 1-a_{i,t} \right)}\le K_{CC}} \right) \ge 1-\epsilon,  \tag{12}
\end{align*}
%
where $K_{CC}$ represents the maximum tolerable number of hydrogen pipeline failures, which can be determined based on the hydrogen leakage process and emergency response capability \cite{chilou2022Hydrogen_leakage}. $\mathbb{P}(\cdot)$ represents the probability, and $\epsilon$ is the confidence level. Constraint (12) ensures that the number of failures within $K_{CC}$ even under the worst-case failure probability distribution.

\subsubsection{Second-Stage Post-disaster Energy Management}

The optimal energy dispatching model is developed to reflect the cost under the disaster, whose object is to minimize the load-shedding cost, i.e.,
\begin{align*}
	g\left(\boldsymbol{x,y,a} \right) =\sum_{j\in N_E}{\sum_{t\in T}{c_{j,t}^{\mathrm{e,ls}}P_{j,t}^{\mathrm{ls}}}}+\sum_{r\in N_H}{\sum_{t\in T}{c_{r,t}^{\mathrm{h,ls}}G_{r,t}^{\mathrm{ls}}}}. \tag{13}
\end{align*}

The operation constraints are as follows:

a. Component failure logical constraints

According to the logical relationship between failures and states of power lines and hydrogen pipelines, the state $u_{ij,t} = u_{ij,t-1}\land(1-a_{ij,t})$, in which $\land$ represents the logical AND.
$u_{i,t}=1$ represents that power line $ij$ is healthy at moment $t$, and $a_{ij,t}=1$ represents that power line $ij$ is destroyed at moment $t$. 
By linearization, the above logic can be formulated as follows:
\begin{gather*}
	u_{ij,t}\le \,\,a_{ij,t},\,\,\tag{14a}
	\\
	u_{ij,t}\le \,\,u_{ij,t-1},\,\,\tag{14b}
	\\
	u_{ij,t}\ge u_{ij,t-1}+a_{ij,t}-1.\tag{14c}
\end{gather*}

b. Grid operation constraints 

Considering the failure of power lines, the operation constraints of the grid can be expressed as follows:
\begin{align*}
	&\sum_{h\in \delta \left( j \right)}{P_{jh,t}}-\sum_{i\in \pi \left( j \right)}{P_{ij,,t}},
	\\
	&=P_{j,t}^{\mathrm{g}}+P_{j,t}^{\mathrm{H2P}}-P_{j,t}^{\mathrm{P2H}}+P_{i,t}^{\mathrm{sub}}-\left( P_{j,t}^{\mathrm{L}}-P_{j,t}^{\mathrm{ls}}, \right)  \tag{15a}
	\\
	&\sum_{h\in \delta \left( j \right)}{Q_{jh,t}}-\sum_{i\in \pi \left( j \right)}Q_{ij,,t}
	\\
	&=Q_{j,t}^{\mathrm{g}}+Q_{j,t}^{\mathrm{H2P}}-Q_{j,t}^{\mathrm{P2H}}-Q_{i,t}^{\mathrm{sub}}-\left( Q_{j,t}^{\mathrm{L}}-Q_{j,t}^{\mathrm{ls}}, \right)  \tag{15b}
\end{align*}
\vspace*{-15pt}
\begin{gather*}
\allowdisplaybreaks
V_{i,t}^{\mathrm{sqr}}-V_{j,t}^{\mathrm{sqr}}\le M\left( 1-u_{ij,t} \right) +\left( R_{ij}P_{ij,t}+X_{ij}Q_{ij,t} \right) /V_0 ,\tag{15c}
\\
V_{i,t}^{\mathrm{sqr}}-V_{j,t}^{\mathrm{sqr}}\ge -M\left( 1-u_{ij,t} \right) +\left( R_{ij}P_{ij,t}+X_{ij}Q_{ij,t} \right) /V_0 ,\tag{15d}
\\
\allowdisplaybreaks
V^{\mathrm{sqr,\min}}_i \leq V_{i,t}^{sqr}\leq V^{\mathrm{sqr,\max}}_i ,\tag{15e}
\\
-u_{ij,t}P_{ij}^{\max}\le P_{ij,t}\le u_{ij,t}P_{ij}^{\max} ,\tag{15f}
\\
\allowdisplaybreaks
-u_{ij,t}Q_{ij}^{\max}\le Q_{ij,t}\le u_{ij,t}Q_{ij}^{\max}. \tag{15g}
\end{gather*}

Constraints (15a)-(15g) are Linear Distflow models of the grid, which are widely used in the scheduling of distribution networks \cite{m.e.1989Linear_distflow}. Constraints (15c)-(15d) is the relation between nodal voltage and branch power flows, where $M$ is a sufficiently large constant and $M(1-u_{ij,t})$ is used to decouple the voltage between the two sides of the faulted power line. 
Constraints (15f)-(15g) formulate the line power constraints under the considered faults.

c. Hydrogen operation constraints

Since the compressors of hydrogen pipelines are effectively protected during construction, the pipeline pressure can meet the operational requirements \cite{wei2021Pipe_model01}. Thus, we use the model proposed in \cite{wei2021Pipe_model01,chengcheng2017Pipemodel03} to characterize the hydrogen distribution network.
\begin{align*}
	&\sum_{m\in \delta \left( n \right)}{F_{mn,t}}-\sum_{o\in \pi \left( n \right)}{F_{no,t}} 
	\\
	&=F_{n,t}^{\mathrm{UT}}\!-\! F_{n,t}^{\mathrm{s+}}\!+\! F_{n,t}^{\mathrm{s-}}\!-\! F_{n,t}^{\mathrm{L}}\!+\!F_{n,t}^{\mathrm{ls}}\!-\!F_{n,t}^{\mathrm{H2P}}\!+\!F_{n,t}^{\mathrm{P2H}}, \tag{16a}
\end{align*}
\vspace*{-23pt}
\begin{gather*}
-u_{mn,t}F_{mn}^{\max}\leq F_{mn,t}\leq u_{mn,t}F_{mn}^{\max}, \tag{16b}
\\
E_{n,t}=E_{n,t-1}+F_{n,t}^{\mathrm{s+}}\eta ^{s+}-F_{n,t}^{\mathrm{s-}}/\eta ^{\mathrm{s-}}, \tag{16c}
\\
0 \leq E_{n,t}\leq E_{n}^{\max},  \tag{146d}
\\
P_{n,t}^{\mathrm{H2P}}=\beta ^{\mathrm{H2P}}F_{n,t}^{\mathrm{H2P}}, \tag{16e}
\\
P_{n,t}^{\mathrm{P2H}}=\beta ^{\mathrm{P2H}}F_{n,t}^{\mathrm{P2H}}. \tag{16f}
\end{gather*}
 
Constraint (16a) represents the nodal flow balance. Constraint (16b) limits the flow capacities considering the hydrogen pipeline failure. Constraint (16d) denotes the variation relationship of energy storages. Constraints (16e)-(16f) represent the energy efficiency of H2P fuel cells and P2H electrolyzers. Note that the upper/lower bounds of electricity and hydrogen devices are omitted to conserve space but are addressed in the solution. 

\section{Solution Methodology}

This section starts by listing the compact formulation in Section III-A. In Section III-B, the HLCC is reformulated into a second-order cone. 
Then, lifted variables are introduced to tackle the intractable decision-dependent second-order moment in Section III-C.
In Section III-D, the problem is reformulated into a two-stage mixed-integer second-order cone programming (MISOCP) problem and finally addressed by the column-and-constraint generation algorithm.
\vspace{-10pt}
\subsection{Compact Formulation}

The proposed two-stage resilience enhancement problem can be expressed as follows.
\begin{align*}
	\underset{\boldsymbol{x}}{\min}\,\,&\underset{p_{\boldsymbol{a}} \left( \boldsymbol{x} \right) \in \mathcal{P}_a}{\max}\,\,\mathbb{E} _{P}\left[ \varphi \left( \boldsymbol{x,a} \right) \right], \tag{17a}
	\\
	\mathrm{s}.\mathrm{t}.\;&A\boldsymbol{x}\le \boldsymbol{b} ,\tag{17b}
	\\
	&\underset{p_a\in \mathcal{P}_a}{\mathrm{inf}}\,\,\mathbb{P}\left\{ \boldsymbol{s}^T\boldsymbol{a}\le K_{CC} \right\} \ge 1-\varepsilon,   \tag{17c}
\end{align*}
where the second-stage problem can be represented as
\begin{align*}
	\varphi \left( \boldsymbol{x,a} \right) =&\underset{\boldsymbol{y}}{\min}\,\,\boldsymbol{h}^T\boldsymbol{y}, \tag{17d}
	\\
	&\;\mathrm{s}.\mathrm{t}.\;B\boldsymbol{x}+C\boldsymbol{y}+D\boldsymbol{a}+E\boldsymbol{u}\le \boldsymbol{d}, \tag{17e}
\end{align*} 

The first-stage decisions $\boldsymbol{x}=\{ \boldsymbol{x}^{\mathrm{H}},\boldsymbol{x}^{\mathrm{E}_0}\}$, and the second-stage decisions $\boldsymbol{y}$ contain all the energy dispatch decisions in (13)-(16).
 $A$ and $\boldsymbol{b}$ are the constant coefficient matrices and vectors corresponding to constraints (11a)-(11c). $\boldsymbol{h}$ is the cost vector in (13). $B, C, D, E$ and $\boldsymbol{d}$ are the constant coefficient matrices and vectors corresponding to constraints (14a)-(16f).
For ease of statement, $\boldsymbol{x}^{\mathrm{H}}$ is denoted by $\boldsymbol{x}$ in the solutions below.
\vspace{-6pt}

\subsection{Reformulation of the HLCC}

Because of the discrete random variables in the HLCC (12), conventional methods based on the conditional value at risk (VaR) and duality face a great computational burden due to the curse of dimensionality.
To this end, a method based on projection and Cantelli's inequality is employed to transfer it into a second-order cone.

Let $\boldsymbol{\xi }=\boldsymbol{s}^T\boldsymbol{a}-\boldsymbol{\mu }\left( \boldsymbol{x} \right)$. 
The following relation can be obtained 
\begin{align*}
\underset{p_a\in \mathcal{P}_a}{\mathrm{inf}}\,\,\mathbb{P}\left\{ \boldsymbol{s}^T\boldsymbol{a}\le K_{CC} \right\} =\underset{p_{\xi}\in \mathcal{P}_{\xi}}{\mathrm{inf}}\,\,\mathbb{P}\left\{ \xi \le K’ \right\},  \tag{18}
\end{align*}
where
\begin{gather*}
\!\!\mathcal{P} _{\xi}:=\left\{ P\in \mathcal{M} _+(\mathcal{A} _{\xi})\left| \begin{array}{c}
	\!\!\left| \mathbb{E} _{P}[\xi ] \right|\le \sqrt{\gamma _1}\sqrt{s^T Q \left( \boldsymbol{x} \right) s}\\
	\mathbb{E} _{P}\left[ \xi ^2 \right] \le \gamma _2\left( s^T Q \left( \boldsymbol{x} \right) s \right)\\
\end{array} \right.\!\!\! \right\} ,\tag{19a}
\\
\mathcal{A} _{\xi}=\left\{ \boldsymbol{s}^T\boldsymbol{a}-\boldsymbol{\mu }\left( \boldsymbol{x} \right) |\forall \boldsymbol{a}\in \mathcal{A} _a \right\}  ,\tag{19b}
\\
K’=K_{CC}-\boldsymbol{s}^T\boldsymbol{\mu }\left( \boldsymbol{x} \right) .\tag{20}
\end{gather*}

Subsequently, by defining sets
\begin{gather*}
\!\!\!\mathcal{P} _0(\boldsymbol{x})\!:=\left\{\! \left( \mu ,\sigma \right) \in \mathbb{R} \!\times\! \mathbb{R} _+\!\left| \begin{array}{c}
	\left| \mu \right|\le \sqrt{\gamma _1}\sqrt{s^TQ \left( \boldsymbol{x} \right) s}\\
	\mu ^2+\sigma ^2\le \gamma _2s^TQ \left( \boldsymbol{x} \right) s\\
\end{array} \right. \!\!\right\}, \tag{21}
\\
\mathcal{P} _1\left( \mu ,\sigma \right) :=\left\{ \mathbb{P} \in \mathcal{M} _+(\mathcal{A} _{\xi})\left| \begin{array}{c}
	\mathbb{E} _P[\xi ]=\mu\\
	\mathbb{E} _P[\xi ^2]=\sigma ^2\\
\end{array} \right. \right\} ,\tag{22}
\end{gather*}
there is
\begin{align*}
\underset{p_a\in \mathcal{P}_a}{\mathrm{inf}}\,\,\mathbb{P}\left\{ \boldsymbol{s}^T\boldsymbol{a}\le K_{CC} \right\} =\underset{p_a\in \mathcal{P}_{\xi}}{\mathrm{inf}}\,\,\mathbb{P}\left\{ \xi \le K’ \right\} 
\\
=\,\,\underset{\left( \mu ,\sigma \right) \in \mathcal{P} _0(\boldsymbol{x})}{\mathrm{inf}}\,\,\underset{\mathbb{P} \in \mathcal{P} _1\left( \mu ,\sigma \right)}{\mathrm{inf}}\mathbb{P}\left\{ \xi \le K' \right\}. \tag{23}
\end{align*}

In this case, the problem is divided into two layers. For the inner layer problem, applying Cantelli's inequality \cite{f.p.1929Cantellis}, there is
\vspace{-10pt}
\begin{align*}
\underset{\mathbb{P} \in \mathcal{P} _1\left( \mu ,\sigma \right)}{\mathrm{inf}}\mathbb{P}\left\{ \xi \le K’ \right\} =\begin{cases}
	\frac{\left( K’-\mu \right) ^2}{\sigma ^2+\left( K’-\mu \right) ^2},\; \mathrm{if}\; \mathrm{K’}\ge \mu\\
	0                   \quad   \qquad \qquad\mathrm{otherwise}\\
\end{cases}\!\!\!\!\!.  \tag{24}
\end{align*}

$K_{CC}$ in the HLCC (12) should ensure solvability, i.e., satisfy $\mathrm{K'}\ge \mu$. Therefore, it can be expressed as
\vspace{-4pt}
\begin{align*}
\underset{p_a\in \mathcal{P}_a}{\mathrm{inf}}\,\,\mathbb{P}\left\{ \boldsymbol{s}^T\boldsymbol{a} \le K_{CC} \right\} &=\,\,\underset{\left( \mu ,\sigma \right) \in \mathcal{P} _0(\boldsymbol{x})}{\mathrm{inf}}\,\frac{\left( K’-\mu \right) ^2}{\sigma ^2+\left( K’-\mu \right) ^2}
\\
&=\underset{\left( \mu ,\sigma \right) \in \mathcal{P} _0(\boldsymbol{x})}{\mathrm{inf}}\frac{1}{\left( \frac{\mathrm{\sigma}}{\mathrm{b}-\mathrm{\mu}} \right) ^2+1}. \tag{25}
\end{align*}
The expression is monotonically decreasing with respect to $ \frac{\mathrm{\sigma}}{\mathrm{b}-\mathrm{\mu}} $. 
Therefore, by using the graphical method to solve $\underset{\left( \mu,\sigma \right) \in \mathcal{P} _0(\boldsymbol{x})}{\mathrm{inf}}-\frac{\mathrm{\sigma}}{\mathrm{b}-\mathrm{\mu} }$ \cite{yiling2018CC_solve}, (25) leads to the following form
\begin{align*}
	&\boldsymbol{s}^T\boldsymbol{\mu} \left( \boldsymbol{x} \right) +\left( \sqrt{\gamma _1}+\sqrt{\left( \frac{1-\epsilon}{\epsilon} \right) \left( \gamma _2-\gamma _1 \right)} \right) \sqrt{\boldsymbol{s}^TQ\left( \boldsymbol{x} \right) \boldsymbol{s}} 
	\\
	&\le K_{CC}\,\,,\mathrm{if}\gamma _1/\gamma _2\le \varepsilon ;\tag{26a}
	\\
	&\boldsymbol{s}^T\boldsymbol{\mu} \left( \boldsymbol{x} \right) +\sqrt{\frac{\gamma _2}{\epsilon}}\sqrt{\boldsymbol{s}^TQ\left( \boldsymbol{x} \right) \boldsymbol{s}}\le K_{CC}\,\,, \mathrm{if} \gamma _1/\gamma _2>\varepsilon .\tag{26b}
\end{align*}

Consequently, the chance constraint (12) is equivalently transformed into second-order cone constraints (26a)-(26b), which can be solved by the optimization solver.
The transformation is equivalent since the projection (18) and the layered analysis (21)-(25) in the transformation do not break the boundary conditions of the chance constraints.


\begin{remark}
	The proposed HLCC and solution method provide a criterion for determining the minimum hardening budget. By replacing the objective function with the minimum hardening budget and retaining only the HLCC (12), the minimum budget required to achieve the desired risk level under varying disaster intensities can be easily computed.
\end{remark}
\vspace{-7pt}
\subsection{Simplification of Ambiguity Sets}
Since the inverse of the second-order moment $Q(\boldsymbol{x})^{-1}$ in (6) cannot be expressed analytically,
the decision-dependent second-order moment based DRO problem is hard to solve. 
Therefore, lifted variables are employed to refine the main cross-moments, recasting the ambiguity set into a lifted partial cross-moment ambiguity set (LPCAS), which still has more statistical information compared with existing methods.
By introducing coefficient vectors $F=\{f_1,... .f_k\}$, the second-row constraint of the ambiguity set (4a) can be approximated as
\vspace{-2pt}
\begin{align*}
	\mathbb{E} _{P}\left[ \left( f_i^{T}(\boldsymbol{a}-\bar{\boldsymbol{\mu}}(\boldsymbol{x})) \right) ^2 \right] \leq \gamma _{2}{f_i}^TQ(\boldsymbol{x})f_i,\,\,\forall i\in \left[ K \right].   \tag{27}
\end{align*}
\vspace{-2pt}

Then, by introducing the lifted variables $w_i$, the above constraint is equivalent to
\vspace{-2pt}
\begin{gather*}
	\mathbb{E} _{P}\left[ w_i \right] \leq \delta _i\left( x \right) \,\,  \forall i\in \left[ K \right] ,\tag{28}
	\\
	\left( f_i^{T}(\boldsymbol{a}-\bar{\boldsymbol{\mu}}(\boldsymbol{x})) \right) ^2\le w_i\,\,,  \delta _i\left( x \right) =\gamma _{2}{f_i}^TQ(\boldsymbol{x})f_i. \tag{29}
\end{gather*}

Meanwhile, the first constraint of (6a) can be approximated as
\vspace{-8pt}
\begin{align*}
	\check{\boldsymbol{\mu}}\left( \boldsymbol{x} \right) \preceq \mathbb{E} _P\left[ \boldsymbol{a} \right] \preceq \hat{\boldsymbol{\mu}}\left( \boldsymbol{x} \right) ,\tag{30}
\end{align*}
where $\check{\boldsymbol{\mu}}\left( \boldsymbol{x} \right)$ and $\hat{\boldsymbol{\mu}}\left( \boldsymbol{x} \right)$ can be derived by identifying the corresponding quantile across each dimension.

As a result, the proposed LPCAS can be obtained:
\begin{align*}
	\mathcal{P}_a^{L} (\boldsymbol{x}):=\left\{ P_{(a,w)}\!\in \!\mathcal{M} _+(\mathcal{A}_a^L )\left| \!\!\begin{array}{c}
		\check{\boldsymbol{\mu}}\left( \boldsymbol{x} \right) \preceq \mathbb{E} _P\left[ \boldsymbol{a} \right] \preceq \hat{\boldsymbol{\mu}}\left( \boldsymbol{x} \right)\\
		\mathbb{E} _P\left[ w_i \right] \le \delta _i\left( x \right) \,\,\forall i\in \left[ K \right]\\
	\end{array} \right. \!\!\! \right\}\!, \tag{31a}
\end{align*}
\vspace{-10pt}
\begin{align*}
	\!\!\!\!\mathcal{A}_a^L \!=\!\left\{\! \left( \boldsymbol{a},\!\boldsymbol{w} \right)\! \left| \!\! \begin{array}{c}
		w_i\ge \left( f_i^{T}(\boldsymbol{a}-\bar{\boldsymbol{\mu}}(\boldsymbol{x})) \right) ^2, \,\forall i\in \left[ K \right]\\
		\boldsymbol{1}^T\boldsymbol{a}\le N^L,\,\boldsymbol{a}\in \{0,1\}^{\left( |I|+|\mathcal{L} |+|N| \right) \cdot T}
	\end{array} \right.\!\!\!\! \right\}\!,\!\!\! \!\!\tag{31b}&&
\end{align*}
where $N^L$ bounds the failure number of power lines and hydrogen pipelines, which can be calibrated based on reliability analyses \cite{sadra2020Distributionally_NK}.

\vspace{-6pt}
\subsection{Solution of the Two-Stage DRO Problem}

This subsection solves the two-stage DRO problem. Once the first-stage decision $\boldsymbol{x}$ is given, based on the strong duality, 
the inner problem
$
\underset{\boldsymbol{p}_{(a,w)}\in \mathcal{P}_a^L}{\max}\,\,\mathbb{E} _{(\boldsymbol{a,w})}\left[ \,\,\varphi \left( \boldsymbol{x},\boldsymbol{a,w} \right) \right]
$
can be transformed into 
\begin{align*}
	&\underset{\boldsymbol{p}_{(a,w)}\in \mathcal{P}_a^L}{\max}\mathbb{E} _{(\boldsymbol{a,w})}\left[\varphi \left( \boldsymbol{x},\boldsymbol{a,w} \right) \right] 
	\\
	&=\!\!\!\!
	\min_{\substack{\boldsymbol{\alpha} \succeq 0,\\
			\boldsymbol{\beta} \succeq 0,\boldsymbol{\gamma } \succeq 0}}\,
\underset{(\boldsymbol{a,w})\in \mathcal{A}'}{\max}\;\varphi  \left( \boldsymbol{x},\boldsymbol{a} \right) 
	 \!+\!\boldsymbol{\alpha}^T\!\!\left( \hat{\boldsymbol{\mu}}\left( \boldsymbol{x} \right) \!-\!\boldsymbol{a} \right) \!-\!\boldsymbol{\beta}^T\!\!\left( \check{\boldsymbol{\mu}}\left( \boldsymbol{x} \right) \!-\!\boldsymbol{a} \right) \!
	 \\
	 &\qquad \qquad \qquad \quad  \quad +\!\boldsymbol{\gamma }^T\left( \boldsymbol{\delta }\left( \boldsymbol{x} \right) -\boldsymbol{w} \right),  \tag{32}
\end{align*}
where $\boldsymbol{\alpha},\boldsymbol{\beta}$ and $\boldsymbol{\gamma}$ are the dual variables of the ambiguity set constraints, respectively. Combined with the first-stage problem, the overall problem can be written as
\begin{align*}
	&\underset{\boldsymbol{x},\boldsymbol{\alpha}\succeq 0,\boldsymbol{\beta}\succeq0,\boldsymbol{\gamma}\succeq 0}{\min}\,\,\boldsymbol{c}^T\boldsymbol{x}+\boldsymbol{\alpha}^T\hat{\boldsymbol{\mu}}\left( \boldsymbol{x} \right) -\boldsymbol{\beta}^T\check{\boldsymbol{\mu}}\left( \boldsymbol{x} \right) +\boldsymbol{\gamma}^T\boldsymbol{\delta }\left( \boldsymbol{x} \right) 
	\\
	&\qquad \qquad +\underset{(\boldsymbol{a},\boldsymbol{w})\in \mathcal{A}_a^L}{\max}\,\underset{\boldsymbol{y}}{\min}\,\boldsymbol{h}^T\boldsymbol{y}-\left( \boldsymbol{\alpha}-\boldsymbol{\beta} \right) ^T\boldsymbol{a}-\boldsymbol{\gamma }^T\boldsymbol{w}\tag{33}
	\\
	&\qquad \qquad\mathrm{s}.\mathrm{t}.{\mathrm{(9a)-(9c),(12a)-(14k),(24).}}
\end{align*}

In problem (33), $\hat{\boldsymbol{\mu}}\left( \boldsymbol{x} \right)$ and $ \check{\boldsymbol{\mu}}\left( \boldsymbol{x} \right) $ are linear about $\boldsymbol{x}$, and $\boldsymbol{\delta}(\boldsymbol{x})$ is bilinear about $\boldsymbol{x}$. Since $\boldsymbol{x}$ is a binary vector, the latter can be transformed into a primary by introducing the auxiliary variable $x^{\mathrm{sqr}}_{ij}=x_ix_j$. At the same time, the product of dual variables $\boldsymbol{\alpha},\boldsymbol{\beta}$ and uncertain variables $\boldsymbol{a}$ can also be linearized via the big-M method.

Consequently, the problem (17) is transformed into a two-stage robust optimization problem, which can be solved using the column-and-constraint generation algorithm. Specifically, the problem (33) is decomposed into the master problem (MP)
	\vspace{-5pt}
\begin{align*}
 &\!\!\!
			\min_{\substack{\boldsymbol{x},\boldsymbol{\alpha}\succeq 0,\\\boldsymbol{\beta}\succeq 0,\boldsymbol{\gamma}\succeq 0}}\;
	\!\!\!\!\!\!\,\,\boldsymbol{c}^T\boldsymbol{x}+\boldsymbol{\alpha}^T\hat{\boldsymbol{\mu}}\left( \boldsymbol{x} \right) -\boldsymbol{\beta}^T\check{\boldsymbol{\mu}}\left( \boldsymbol{x} \right)  +\boldsymbol{\gamma }^T\boldsymbol{\delta }\left(\boldsymbol{x} \right) +L, \!\!\!\tag{34a}&&
	\\
	& \quad\mathrm{s.t.}\;\;A\boldsymbol{x}\le\boldsymbol{b}, \tag{34b}
	\\
	& \quad \qquad L\ge \boldsymbol{h}^T\boldsymbol{y}-\left( \boldsymbol{\alpha}-\boldsymbol{\beta} \right) ^T\boldsymbol{a}-\boldsymbol{\gamma }^T\boldsymbol{w},\tag{34c}
	\\
	&\quad \qquad B\boldsymbol{y}\le \boldsymbol{d}\,\,;C\boldsymbol{x}+D\boldsymbol{y}\le \boldsymbol{e},\,\,\tag{34d}
	\\
	&\quad \qquad E\boldsymbol{y}+F\boldsymbol{a}^{(k)}\le \boldsymbol{f}\,\,\,\,\forall \boldsymbol{a}^{(k)}\in \mathcal{A} ^{(k)}, \tag{34e}
	\\
	&\quad \qquad \boldsymbol{c}^T\boldsymbol{x}\!+\!\boldsymbol{\alpha}^T\hat{\boldsymbol{\mu}}\left( \boldsymbol{x} \right)\!-\!\boldsymbol{\beta}^T\check{\boldsymbol{\mu}}\left( \boldsymbol{x} \right)  \!+\!\boldsymbol{\gamma }^T\boldsymbol{\delta }\left(\boldsymbol{x} \right) \!+\! L \geq 0,  \tag{34f}
\end{align*}
	\vspace{-5pt}
and the subproblem (SP)
\begin{align*}
	&\underset{(\boldsymbol{a},\boldsymbol{w})\in \mathcal{A} _{a}^{L}}{\max}\,\underset{\boldsymbol{y}}{\min}\;\boldsymbol{h}^T\boldsymbol{y}-\left( \boldsymbol{\alpha }-\boldsymbol{\beta } \right) ^T\boldsymbol{a}-\boldsymbol{\gamma }^T\boldsymbol{w},\tag{35a}&&
\\
&\quad \; \mathrm{s}.\mathrm{t}.\left( 11\mathrm{a} \right) -\left( 11\mathrm{c} \right) ,(14\mathrm{a}-16\mathrm{f}),(26).\tag{35b}
\end{align*}
	\vspace{-12pt}
\begin{algorithm}[t]
	\caption{Column-and-constraint Generation Algorithm for the Transformed Two-stage Model.}  
	\label{algorithm2}
	\LinesNumbered  
	\textbf{Initialize: } Set $\mathrm{UB}=\infty$, $\mathrm{LB}=0$, $k=0$,$\mathcal{A}_{(0)}=\emptyset$;
	
	\While{$\left|\frac{\mathrm{UB}-\mathrm{LB}}{\mathrm{UB}}\right| > \epsilon $}{
		\textbf{Step1: }Solve the MP (32) with set $\mathcal{A} _{(k)}$. Derive the optimal solution ($\boldsymbol{x}_{(k)}^*,\boldsymbol{\alpha}_{(k)}^*,\boldsymbol{\beta}_{(k)}^*,\boldsymbol{\gamma}_{(k)}^*$), the optimal value $MP_{(k)}^*$, and the value $L_{(k)}^*$.\\
		\textbf{Step2: }Update $\mathrm{LB}=MP_{(k)}^*$.\\
		\textbf{Step3: }Solve the SP$'$ (33) with the first stage decision ($\boldsymbol{x}_{(k)}^*,\boldsymbol{\alpha}_{(k)}^*,\boldsymbol{\beta}_{(k)}^*,\boldsymbol{\gamma}_{(k)}^*$). Derive an optimal solution $\mathrm{UB}=\boldsymbol{a}_{(k)}$ and the optimal value $SP_{(k)}^*$.\\
		\textbf{Step4: }Update $\mathrm{UB}=\min\{\mathrm{UB},{MP_{(k)}^*-L_{(k)}^*+SP_{(k)}^*}\}$.\\
		\textbf{Step5: }Update $\mathcal{A} _{(k+1)} = \mathcal{A} _{(k)} \cup  \{ \boldsymbol{a}_{(k)}\}$. Let $k=k+1$.\\
	}
	\textbf{Return } UB, $\boldsymbol{x}_{(k)}^*$.

\end{algorithm}

Based on the strong duality, SP can be transformed into SP$'$
\begin{align*}
	\!\!&\max_{\substack{(\boldsymbol{a},\boldsymbol{w})\in \mathcal{A}_a^L ,\\ \boldsymbol{\tau },\boldsymbol{\upsilon },\boldsymbol{\omega }}}
	\!\!(\boldsymbol{\beta}-\boldsymbol{\alpha })^T\boldsymbol{a}-\boldsymbol{\gamma }^T\boldsymbol{w}-\boldsymbol{\tau} ^T\boldsymbol{d}+\upsilon ^T\left( C\boldsymbol{x}-\boldsymbol{e} \right)
	\\
	& \qquad  \quad +\boldsymbol{\omega}^T\left( F\boldsymbol{a}-\boldsymbol{f} \right) ,\tag{36a}
	\\
	&\quad  \;\; \mathrm{s}.\mathrm{t}.\;\;\boldsymbol{h}+B^T\boldsymbol{\tau }+D^T\boldsymbol{\upsilon }+E^T\boldsymbol{w}=0, \tag{36b}
	\\
	&\qquad \quad\;\,\boldsymbol{\tau }\ge 0,\boldsymbol{\upsilon }\ge 0,\boldsymbol{\omega }\ge 0 .\tag{36c}
\end{align*}

As shown in Algorithm \ref{algorithm2}, the column-and-constraint generation algorithm alternately solves the MP and the SP$'$. By iteratively adding cut planes to the MP while updating the upper and lower bounds, the process will converge.
After that, the optimal resilience enhancement policy can be obtained.

\section{Case Studies}
\subsection{Parameter Settings}
\begin{figure}[!t]
	\centering
	\includegraphics[width=3.5in]{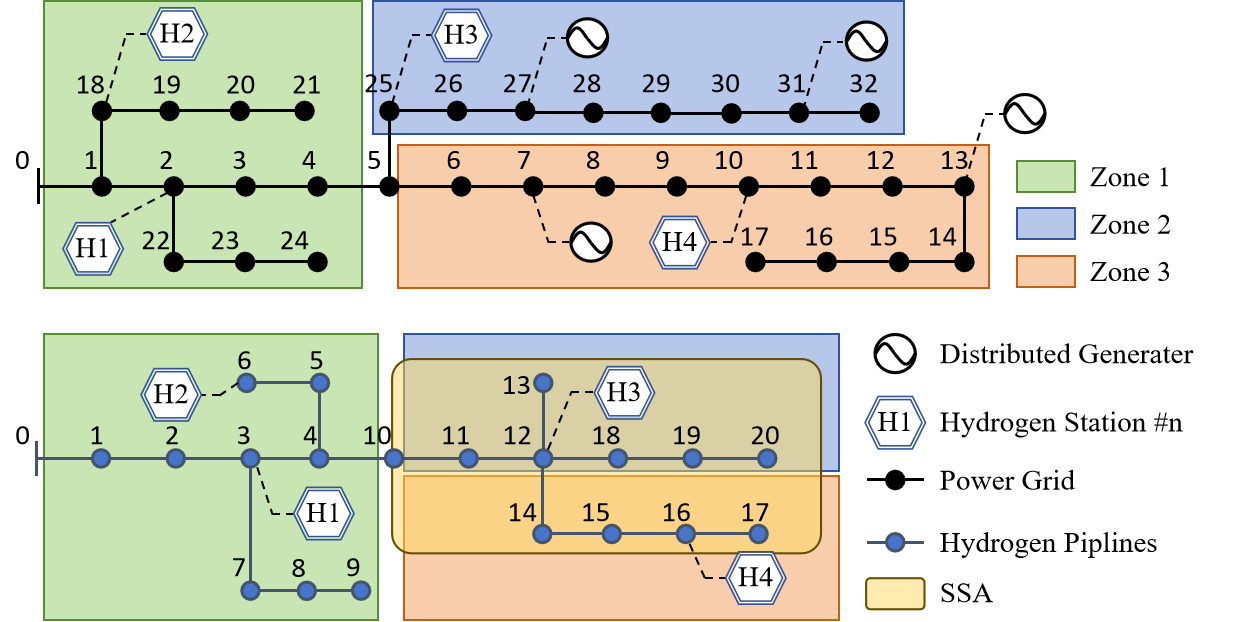}
	\caption{Electricity-hydrogen distribution system with SSA. 
	}
	\label{fig_case}
	\vspace{-4pt}
\end{figure}

\begin{table*}[t]
	\centering
	\caption{Case setup}
	\label{tab1_disaster}
	\begin{tabular}{c c c c c}
		\toprule
		Case ID & System Components & \makecell{Disaster \\ Level} & \makecell{Expectations of\\ Maximum Wind $(\mathrm{m/s})$} & \makecell{Expectations of\\ Maximum Rainfall $(\mathrm{mm/h})$}\\
		\midrule
		1 & Proposed model & 1 & [35,40] & [11,14] \\ 
		2 & Proposed model & 2 & [40,45] & [14,17] \\ 
		3 & Proposed model & 3 & [45,50] & [17,20] \\ 
		4 & Proposed model & 4 & [50,55] & [20,22] \\ 
		5 & \makecell{Without hydrogen storages}  & 3 & [45,50] & [17,20] \\
		6 & \makecell{Without hydrogen fuel cells and electrolyzers}  & 3 & [45,50] & [17,20] \\
		\bottomrule
	\end{tabular}
\end{table*}

The proposed method in this paper is validated on a modified model based on the IEEE-33 distribution network coupled with a 21-node hydrogen pipeline network through distributed hydrogen stations. As shown in Fig. \ref{fig_case}, both the electricity and hydrogen networks are divided into three zones based on their geographical characters.
There are four DGs and four hydrogen stations which contain fuel cells, electrolyzers, and hydrogen storage.
The hydrogen pipelines in the yellow area are within the SSA. 
The voltage standard of the distribution network is 12.66 KV, and the voltage limits are set to [0.9,1.1] p.u. 
To simulate hourly load fluctuations during the disaster, the power and hydrogen loads for each hour are obtained by multiplying the standard load by a random factor [0.8,1.2].

\textcolor{black}{The total hardening budget is set to be $5.5 *10^5$ \$. 
	The cost of hardening for the power line and hydrogen pipeline is set to 20000 \$/km \cite{yujia2023DDUDRO02} and 37500 \$/km, respectively. }
Assume that the distance between two consecutive poles is 50 meters, and the length of hydrogen pipelines in the fragility model (6) is 200 meters.
The fragility models of the lines are obtained by fitting \cite{jeffreya.2021Fragility} and \cite{chuan2018Back05}, respectively.
The impact of hardening references the \cite{mingyu2019Pipemodel02}. 
Based on the energy ratio, the basic load shedding penalty of electricity and hydrogen is set to 15 \$/kwh \cite{xu2019Datacost02} and 100 $\mathrm{\$/m^3}$, respectively. 
	Besides, to reflect the severity differences of load shedding among different nodes,
	power nodes $\{2,18,26\}$ and hydrogen nodes $\{3,13,18\}$ are set as critical nodes. The load-shedding weights of the non-critical nodes are randomly generated from interval $[1,5]$, and the weight of critical loads is set to 50 \cite{yujia2023DDUDRO02}.
To reflect different disaster intensities and energy equipment configurations, six cases are set up as shown in Table \ref{tab1_disaster}, with $t=12$.
Besides, to simulate the impact of geographic factors such as typhoon movement on the intensities of the disaster, the intensities are set as $[0.5,1]$ increasing and $[1,0.5]$ decreasing of max intensities during $t=1\sim6$ and $t=7\sim12$ respectively. The variances of wind speed and rainfall were set to 4 and 9, respectively. 

In SSA, $K_{CC}$ is set to 1 since the hydrogen leakage is severe.
The detailed network data with length of lines and device parameters can be found in \cite{case_data}.

The computation is performed in Python 3.11 and solved by Gurobi 11.0, on an Intel Core i9-13900 CPU with 64 GB RAM PC. 
\vspace{-5pt}
\subsection{Results of the Proposed Model}
To validate the effectiveness of the proposed algorithm and also verify the effects of hydrogen-electricity coupling, hardening decisions, storage allocation decisions, and worst expected load-shedding costs (WELSCs) under Case 1-6  indicated in Table I are demonstrated in Table \ref{tab1_result}.
By comparing Cases 1-4, it can be seen that the WELSC increases as the disaster level increases. This is due to the fact that greater-intensity disasters increase the probability of line failure, thus reducing the energy supply abilities of the EHDN.
Meanwhile, the number of hydrogen pipeline hardening also gradually increases to satisfy the HLCC.

\begin{figure*}[!t]
	\centering
	\subfloat[]{\includegraphics[width=3.5in]{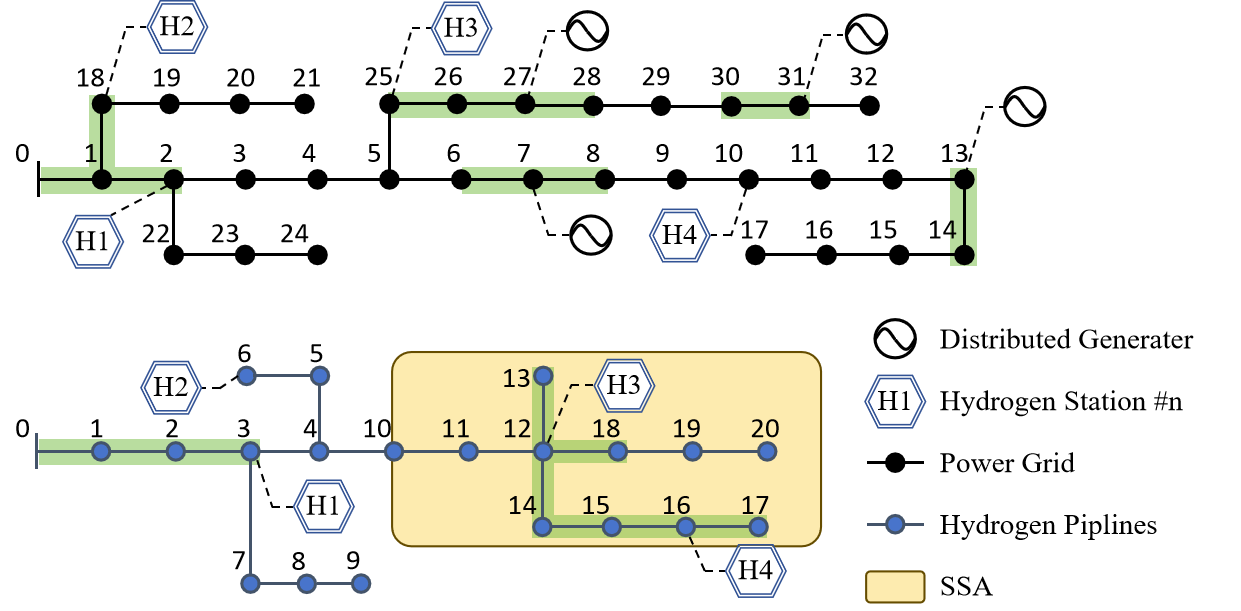}
		\label{fig_result1}}
	\hfil
	\subfloat[]{\includegraphics[width=3.5in]{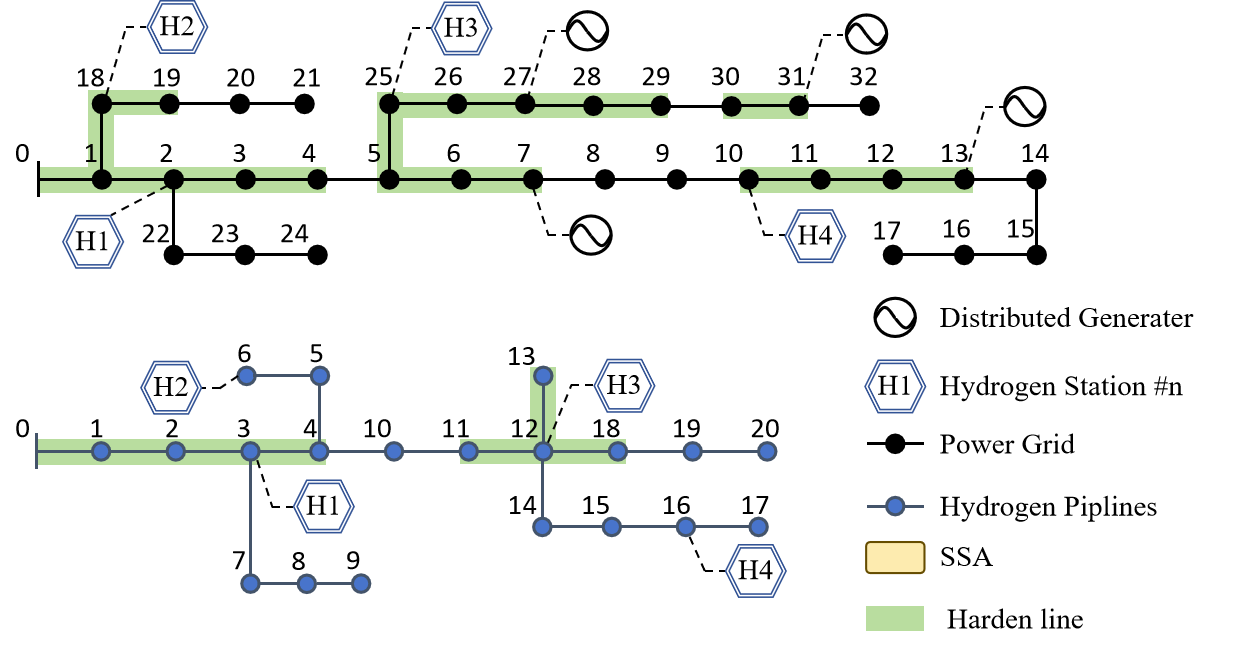}%
		\label{fig_result2}}
	\caption{Hardening decision under Case 3 (a) with the HLCC; (b) without the HLCC.}
	\label{fig_sim}
	\vspace{-10pt}
\end{figure*}
\begin{table}[t!]
	\centering
	\caption{Results of proposed method in each case}
	\label{tab1_result}
	\begin{tabular}{c c c c c}
		\toprule
		\multirow{3}{*}[-3pt]{Case} & \multirow{3}{*}[-3pt]{\makecell{Hydrogen Storages\\Allocation ($\mathrm{m^3}$) }}& \multicolumn{2}{c}{Hardening Number}&\multirow{3}{*}[-3pt]{WELSCs (\$)}\\ 	\cmidrule(r){3-4}
		&&power lines&\makecell{hydrogen\\pipelines}&\\\midrule
		1  &  \{150, 0, 150, 100\} & 15 & 6 & $1.44*10^6$\\  
		2  &  \{150, 0, 150, 100\} & 13 & 7 & $2.07*10^6$\\ 
		3  &  \{150, 0, 150, 100\}& 10 & 9& $2.54*10^6$\\
		4 & \{150, 0, 150, 100\} & 6 & 11& $2.90*10^6$\\
		5 & No storages & 7 & 11 & $4.04*10^6$\\
		6 & \{150, 0, 150, 100\} & 8 & 10 & $3.13*10^6$\\
		\bottomrule
	\end{tabular}
	\vspace{-10pt}
\end{table}

Besides, in Table II, the WELSCs in cases without hydrogen storage (Case 5) and without hydrogen fuel cells and electrolyzers (Case 6) are significantly higher than in Case 3. 
This validates the enhancement of energy supply capacity and flexibility by hydrogen storage and P2H/H2P conversions, especially in disaster situations.

To more intuitively demonstrate the hardening decisions, Fig. \ref{fig_result1} shows the exact hardening choice under Case 3. 
To minimize the energy load shedding, it can be seen that hardened lines mainly concentrated in root areas of the networks (like 0-2, 1-18 in the power network, and 0-3 in the hydrogen network), and areas around hydrogen stations or distributed generators (like 25-28, 6-8 in the power network and 12-18 in the hydrogen network).
This is because the hardening of these areas can best enhance the energy supply capacity of the system and thus reduce the WELSCs.
Besides, many hydrogen pipelines in SSA are hardened, safeguarding hydrogen leakage risk in this area.
The above proves the validity of the proposed model and method.

\subsection{Comparative Experiments on the HLCC}
To validate the effectiveness of the HLCC,
under the disaster level 1-4 indicated in Table \ref{tab1_disaster}, the hardening numbers in SSA (HN-SSA), 95\% Value-at-risk of failure in SSA (VaR-SSA), and WELSCs are compared both with and without the HLCC (10). 
By randomly sampling 1000 scenarios, VaR-SSA (calculated using the 5\% worst case) is computed. 

The results are shown in Table \ref{tab2_CC}. 
Under disaster levels 2, 3, and 4, the HN-SSA are larger when containing the HLCC. 
Meanwhile, the VaR-SSA are no larger than $K_{CC}$ with the HLCC, meaning that the risk of hydrogen pipelines is effectively limited.
Besides, under disaster level 1, the intensity is sufficiently small that hardening is unnecessary to satisfy the HLCC.
Thus the results are the same with or without the HLCC.
Naturally, WELSCs are a bit higher when containing the HLCC, but as mentioned earlier, hydrogen safety is necessary, which is much more critical than energy supply.

Fig. \ref{fig_result2} labels the hardening decisions without the HLCC.
In contrast to Fig. \ref{fig_result1}, fewer hydrogen pipelines are hardened within the SSA. Besides, more power lines are hardened due to the excess budget, explaining the WELSC differences in Table \ref{tab2_CC}.


%
%

\begin{table}[t]
	\centering
	\caption{Compare experiments results about the HLCC}
	\label{tab2_CC}
	\begin{tabular}{c c c c c}
		\toprule
		\makecell{Disaster \\ level} & \makecell{Whether contain\\ the HLCC}  & HN-SSA & VaR-SSA &   WELSC (\$)  \\ \midrule
		\multirow{2}{*}[0pt]{1} &  Yes & 2 &  1  & $1.44*10^6$\\  \cline{2-5}
		&  No & 2 &  1  & $1.44*10^6$ \\ \hline
		\multirow{2}{*}[0pt]{2}  &  Yes & 3  & 1  & $2.07*10^6$\\ \cline{2-5}
		& No & 2 & 2  & $1.98*10^6$ \\ \hline
		\multirow{2}{*}[0pt]{3} & Yes & 6 &1  & $2.54*10^6$ \\ \cline{2-5}
		& No & 3 & 2 & $2.37*10^6$ \\ \hline
		\multirow{2}{*}[0pt]{4} & Yes & 7 & 1  & $2.90*10^6$ \\ \cline{2-5}
		& No & 3 & 3 & $2.66*10^6$\\
		\bottomrule
	\end{tabular}
	\vspace{-10pt}
\end{table}

%
%
%
%

\subsection{Compare Experiments on Ambiguity Sets}
\begin{figure}[!t]
	\centering
	\includegraphics[width=3.5in]{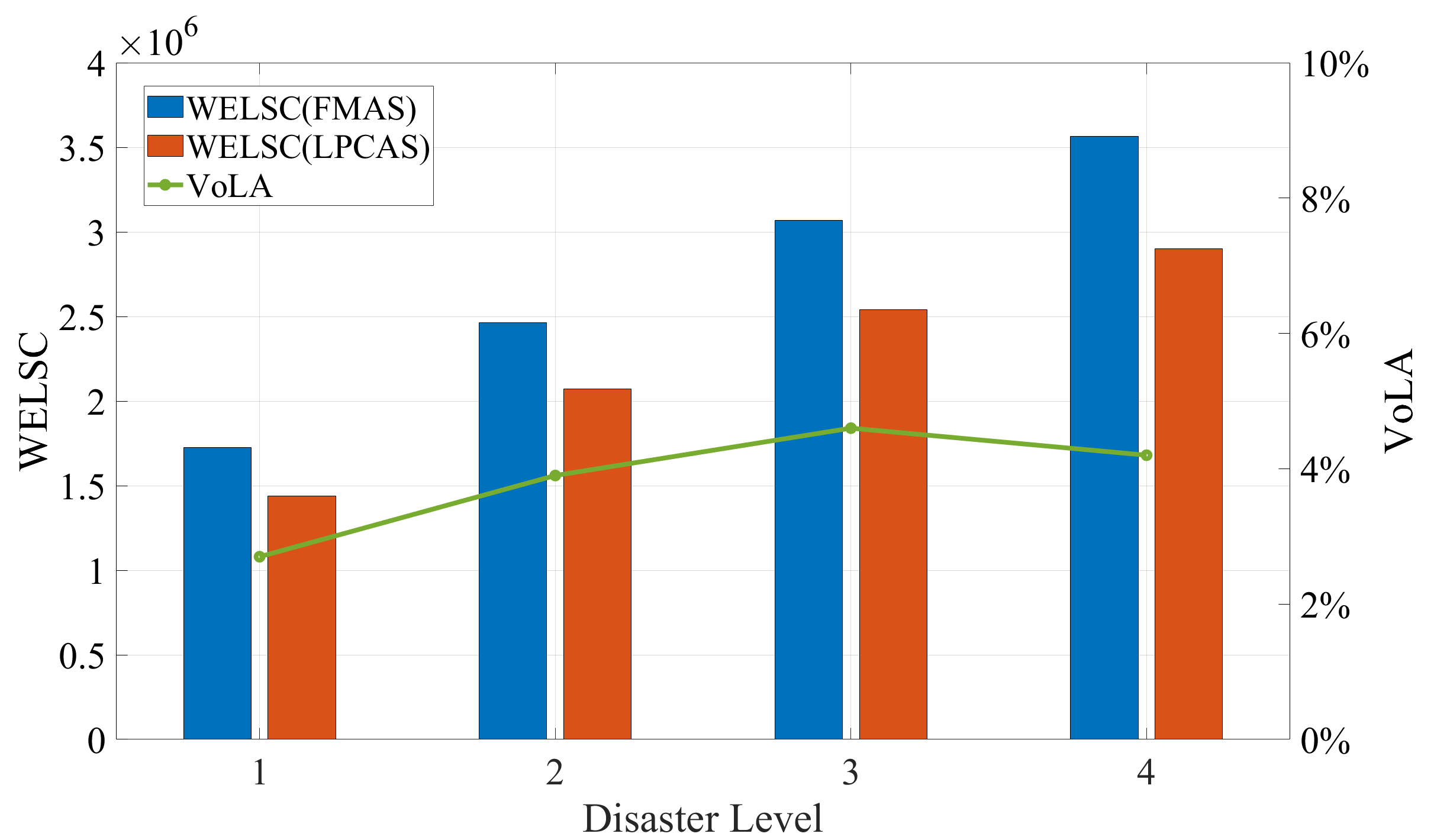}
	\caption{The WELSCs and VoLA under different ambiguity sets.}
	\label{fig_test2}
	\vspace{-10pt}
\end{figure}

To demonstrate the benefits of LPCAS, which incorporates the main cross-moments of the second-order moment, comparative experiments about the ambiguity set are developed under the disaster level 1-4 in Table \ref{tab1_disaster}.
By retaining only the first-row constraint in (29), the decision-dependent first-order moment based ambiguity set (FMAS) is constructed, which is commonly used in existing works.
The WELSCs based on each set are shown in Fig. \ref{fig_test2}.
It can be seen that the WELSCs are lower when based on the LPCAS than the FMAS.
It is because the LPCAS provides a more comprehensive estimate of uncertainties and reduces over-conservation, so that the estimation of the worst-case scenario is also more rigorous.
Of course, the difference in WELSCs can not directly reflect the decision performance since the worst probability distributions are different.

To intuitively quantify the impact, the metrics proposed by \cite{hamed2019Vola} are introduced. 
Specifically, the value of the lifted ambiguity set (VoLA) is defined as follow:
$$
\mathrm{VoLA}=\frac{F_{}^{\mathrm{LPCAS}}\left( \boldsymbol{x}^{\mathrm{FMAS}} \right) -F_{}^{\mathrm{LPCAS}}\left( \boldsymbol{x}^{\mathrm{LPCAS}} \right)}{F_{}^{\mathrm{LPCAS}}\left( \boldsymbol{x}^{\mathrm{LPCAS}} \right)}
$$
where $F^{\mathrm{LPCAS}}(\cdot)$ represents the WELSC based on the LPCAS. $\boldsymbol{x}^{\mathrm{FMAS}}$ and $\boldsymbol{x}^{\mathrm{LPCM}}$ represent the optimal decision obtained according to the FMAS and LPCAS, respectively.
The results in Fig. \ref{fig_test2} show that LPCAS brings about 3\% to 4.5\% performance improvement. 
This is because LPCAS incorporates more information about failure probability distributions, 
which leads to more effective hardening decisions to enhance resilience under limited budgets. 
\vspace{-10pt}
\subsection{The Minimum Hardening Budgets for the HLCC}
As described in Section III-B, the processing of the HLCC also provides a method to get the minimum hardening budget to limit the danger of hydrogen pipelines in the SSA.
This subsection provides the results under different disaster levels in the HLCC (12),
as shown in Table \ref{table_budget}.
It can be seen that the minimum hardening budget increases monotonically with increasing disaster levels.
Also, under disaster level 1, hardening in the SSA is not necessary, so the minimum budget is zero. This is consistent with the conclusions in Section IV-D.


\begin{table}[t]
	\centering
	\caption{Minimum hardening budget through the HLC}
	\label{table_budget}
	\begin{tabular}{c | c| c| c| c}
		\toprule
		Disaster Level & 1 & 2 & 3 & 4 \\
		\midrule
		Minimum budget (\$) & $0$ & $1.03*10^4$ & $2.25*10^4$ & $2.82*10^4$  \\ 
		\bottomrule
	\end{tabular}
\end{table}

	\vspace{-5pt}
%

	\subsection{Large-Scale Experiment}
	To validate the computational efficiency of the proposed method, a large-scale experiment is carried out based on PG\&E 35KV 69-bus power distribution network and 34-node hydrogen distribution network modified from \cite{li201969Dian}.

\begin{figure}[!t]
	\centering
	\includegraphics[width=3.6in]{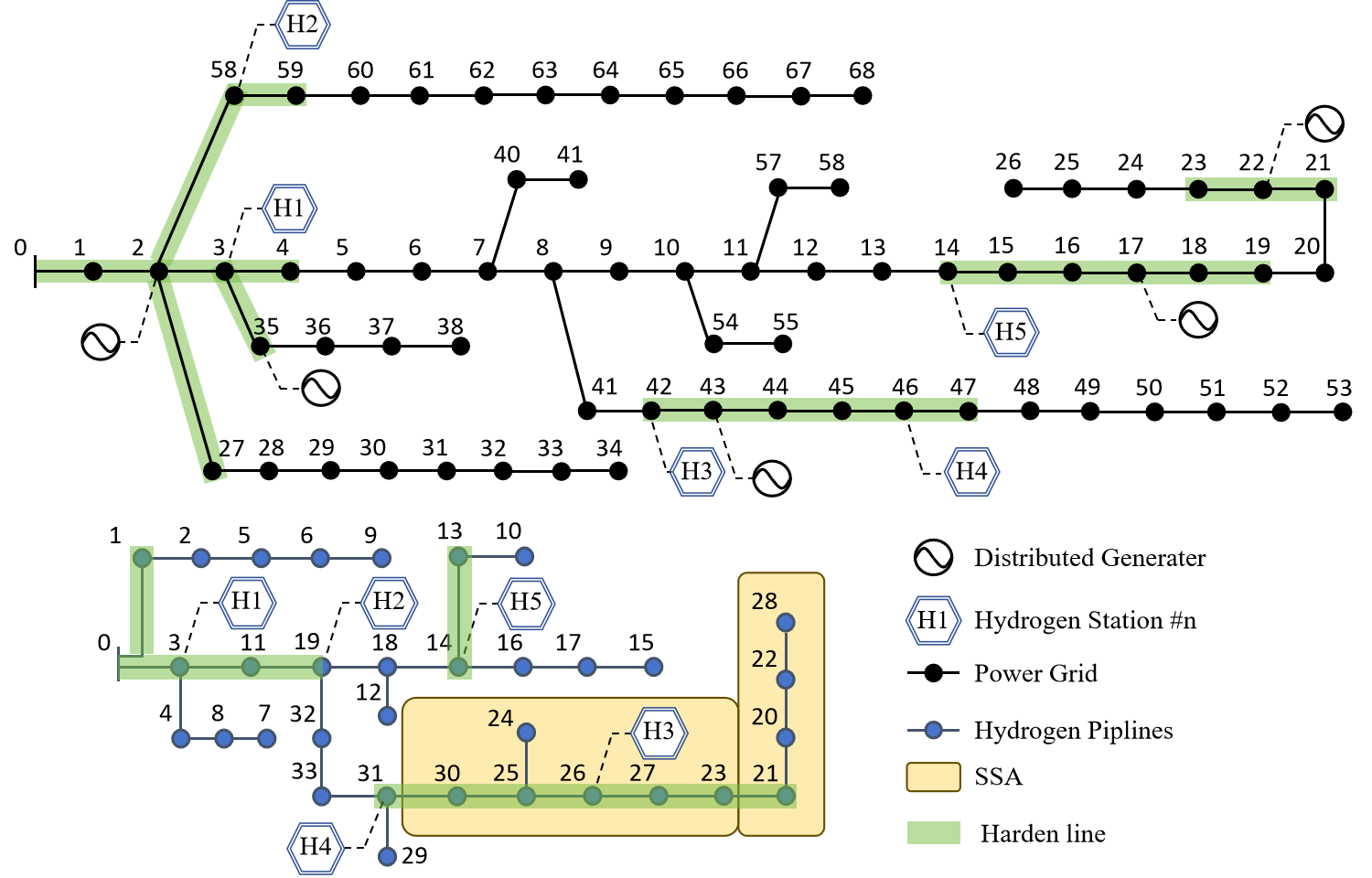}
	\caption{The topology of the PG\&E 69-bus power with 34-node hydrogen distribution network, and hardening results.
	}
	\label{fig_big}
	\vspace{-4pt}
\end{figure}

	The network topology is shown in Fig. \ref{fig_big}. The system contains five distributed generators and five hydrogen stations. 
	The hardening budget is set to $8.5*10^5$ \$. Power nodes \{3,27,46,59\} and hydrogen nodes \{2,13,19,26\} are set as critical nodes. In SSA, the $K_{CC}$ is also set to 1.
	Other system parameters, including disaster geographic zoning, line lengths, and equipment capacity can be found in \cite{case_data}.

	The algorithm converges in four hours and ten minutes. The value of WELSC is $6.38*10^7$ \$. 
	The hardening results are shown in Fig. \ref{fig_big}. 
	It can be seen that hardening centers around root nodes, distributed generators, hydrogen stations, critical nodes, and within the SSA. 
	This validates the effectiveness of the algorithm with the goal of minimizing the worst expected load-shedding cost.
	In addition, considering that extreme weather can usually be sensed and predicted several days in advance, the calculation time of the proposed method is acceptable.

\section{Conclusion}
%
%
%
Facing the threat of extreme weather events to electricity-hydrogen distribution networks (EHDNs), this paper proposes an optimal defensive hardening strategy that not only minimizes load shedding but, more importantly, controls the risk of hydrogen leakage. This contributes to the wider adoption of hydrogen.
Specifically, first, this paper establishes a more accurate failure uncertainty model that simultaneously considers the effects of hardening and disaster intensity correlation. Subsequently, an optimal hardening decision framework is developed based on a two-stage distributionally robust optimization (DRO) approach incorporating a hydrogen leakage chance constraint (HLCC). Efficient solution methods are then proposed to address challenges introduced by decision-dependent uncertainties (DDUs).
Case studies on both small and large scales demonstrate that the proposed method effectively controls hydrogen leakage risk through hardening. Furthermore, the improved failure uncertainty characterization reduces load-shedding costs.

\bibliographystyle{IEEEtran}
\bibliography{paper_1_new_offline}

%
%
%
%
%
%
%
%
%
%

\newpage

 
\vspace{11pt}

%
%

\vfill

\end{document}